\documentclass{article}[12pt]

\usepackage{latexsym}

\begin{document}
\title{Solvable Dynamical Systems in the Plane with Polynomial Interactions}

\author{Francesco Calogero$^{a,b}$\thanks{e-mail: francesco.calogero@roma1.infn.it}
\thanks{e-mail: francesco.calogero@uniroma1.it}
 , Farrin Payandeh$^c$\thanks{e-mail: farrinpayandeh@yahoo.com}
 \thanks{e-mail: f$\_$payandeh@pnu.ac.ir}}

\maketitle   \centerline{\it $^{a}$Physics Department, University of
Rome "La Sapienza", Rome, Italy}

\maketitle   \centerline{\it $^{b}$INFN, Sezione di Roma 1}

\maketitle

\maketitle   \centerline{\it $^{c}$Department of Physics, Payame
Noor University, PO BOX 19395-3697 Tehran, Iran}

\maketitle

\begin{abstract}
In this paper we report a few examples of \textit{algebraically
solvable} dynamical systems characterized by $2$ coupled Ordinary
Differential Equations which read as follows:
$$
\dot{x}_{n}=P^{\left( n\right) }\left( x_{1},x_{2}\right)
~,~~~n=1,2~,
$$
with $P^{\left( n\right) }\left( x_{1},x_{2}\right) $ specific \textit{%
polynomials} of relatively low degree in the $2$ dependent variables $%
x_{1}\equiv x_{1}\left( t\right) $ and $x_{2}\equiv x_{2}\left(
t\right) $. These findings are obtained via a new twist of a recent
technique to identify dynamical systems \textit{solvable by
algebraic operations}, themselves explicitly identified as
corresponding to the time evolutions of the \textit{zeros} of
polynomials the \textit{coefficients} of which evolve according to
\textit{algebraically solvable} (systems of) evolution equations.

\end{abstract}

\section{Introduction}

It has been recently noted \cite{C2016} that, if the quantities
$x_{n}\left( t\right) $ respectively $y_{m}\left( t\right) $ denote
the $N$ \textit{zeros} respectively the $N$ \textit{coefficients} of
a \textit{generic}
time-dependent monic polynomial $p_{N}\left( z;t\right) $ of degree $N$,%
\begin{equation}
p_{N}\left( z;t\right) =z^{N}+\sum_{m=1}^{N}\left[ y_{m}\left(
t\right) ~z^{N-m}\right] =\prod_{n=1}^{N}\left[ z-x_{n}\left(
t\right) \right] ~, \label{PolN}
\end{equation}%
there hold the following \emph{identities} relating the time
evolution of these quantities:
\begin{equation}
\dot{x}_{n}=-\left[ \prod_{\ell =1,~\ell \neq n}^{N}\left(
x_{n}-x_{\ell }\right) \right] ^{-1}\sum_{m=1}^{N}\left[
\dot{y}_{m}~\left( x_{n}\right) ^{N-m}\right] ~,~~~n=1,2,...,N~.
\label{xnymdot}
\end{equation}

\textbf{Notation 1.1}. Hereafter all quantities are \textit{a
priori} assumed to be \textit{complex} numbers, with the following
exceptions: \textit{indices} such as $n$, $m$, take positive integer
values (over ranges specified on a case-by-case basis: indeed, in
most of this paper the range is limited just to the $2$ values $1$
and $2$ for $n$, and to $1$ and $2$ or $1$, $2$ and $3$ for $m$);
while the independent variable $t$ ("time") is \textit{real} and it
is generally assumed to run from $0$ to $+\infty $. The
$t$-dependence of time-dependent variables such as $x_{n}\left(
t\right) $ and $y_{m}\left( t\right) $ is often not
\textit{explicitly} displayed (even, inconsistently, in the same
formula: of course when this is unlikely to cause
misunderstandings); and superimposed dots on these variables denote
of course time-differentiations, $\dot{x}_{n}\equiv dx_{n}\left(
t\right) /dt,$ $\dot{y}_{m}\equiv dy_{m}\left( t\right) /dt$. It is
of course not
excluded that \textit{complex} numbers take \textit{real} or \textit{%
imaginary} values, as indicated below on a case-by-case basis:
indeed the words "in the plane" in the title of this paper refer to
the standard case in which the two coordinates $x_{1}\left( t\right)
$ and $x_{2}\left( t\right) $ are interpreted as the $2$
\textit{real} coordinates of a\ point moving in the Cartesian
$x_{1}x_{2}$-plane, or as the $2$ \textit{complex} coordinates of
$2$ points moving in the \textit{complex} plane (or, equivalently,
of $2$ \textit{real} two-vectors moving in a plane: see below).
$\Box$

Analogous formulas to (\ref{xnymdot}) also exist for higher
time-derivatives
\cite{BC2016} \cite{BrC2016}, and via such formulas many new \textit{%
algebraically solvable} dynamical systems have been recently
identified and discussed, especially dynamical systems characterized
by second-order Ordinary Differential Equations (ODEs) of Newtonian
type\ ("accelerations equal forces"): for an overview see
\cite{C2018} and references therein. But
in this paper our treatment is confined to systems involving \textit{%
first-order} time-derivatives.

In this paper we moreover confine attention to the very simplest
such systems: characterized by \textit{first-order} Ordinary
Differential Equations involving \emph{only} $2$ dependent
variables. Let us tersely review here---in this very simple
context---how this approach works.

Systems of \textit{algebraically solvable} first-order ODEs for the zeros $%
x_{n}\left( t\right) $ are obtained from the identities
(\ref{xnymdot}) by assuming that the $N$ coefficients $y_{m}\left(
t\right) $ satisfy themselves an \textit{algebraically solvable}
system of first-order ODEs. Note that in the very simple case with
$N=2$ the equations (\ref{xnymdot})
read simply as follows:%
\begin{equation}
\dot{x}_{n}=\left( -1\right) ^{n}\left( \frac{x_{n}~\dot{y}_{1}+\dot{y}_{2}}{%
x_{1}-x_{2}}\right) ~,~~~n=1,2~.  \label{xnymdot12}
\end{equation}%
Now assume that the system of $2$ ODEs

\begin{equation}
\dot{y}_{1}=f_{1}\left( y_{1},y_{2}\right)
~,~~~\dot{y}_{2}=f_{2}\left( y_{1},y_{2}\right) ~,  \label{ymdot12}
\end{equation}%
be \textit{algebraically solvable} (of course, for an appropriate
assignment of the $2$ functions $f_{1}\left( y_{1},y_{2}\right) $
and $f_{2}\left(
y_{1},y_{2}\right) $). Then the system%
\begin{equation}
\dot{x}_{n}=\left( -1\right) ^{n}\left[ \frac{x_{n}~f_{1}\left(
-x_{1}-x_{2},x_{1}x_{2}\right) +f_{2}\left( -x_{1}-x_{2},x_{1}x_{2}\right) }{%
x_{1}-x_{2}}\right] ~,~~~n=1,2  \label{xndot12}
\end{equation}%
is as well \textit{algebraically solvable}, because it clearly
corresponds to (\ref{xnymdot12}) via the $2$ identities

\begin{equation}
y_{1}\left( t\right) =-\left[ x_{1}\left( t\right) +x_{2}\left( t\right) %
\right] ~,~~~y_{2}\left( t\right) =x_{1}\left( t\right) x_{2}\left(
t\right) \label{y12Ex1}
\end{equation}%
clearly associated to the polynomial (\ref{PolN}) with $N=2$,%
\begin{equation}
p_{2}\left( z;t\right) =z^{2}+y_{1}\left( t\right) z+y_{2}=\left[
z-x_{1}\left( t\right) \right] \left[ z-x_{2}\left( t\right) \right]
~. \label{p2zt}
\end{equation}%
Indeed the solution of its initial-values problem---to compute
$x_{1}\left( t\right) $ and $x_{2}\left( t\right) $ via
(\ref{xnymdot12}) from the
assigned initial data $x_{1}\left( 0\right) $ and $x_{2}\left( 0\right) $%
---can be achieved via the following $3$ steps: \textit{(i)} from the
initial data $x_{1}\left( 0\right) $ and $x_{2}\left( 0\right) $
compute the corresponding initial data $y_{1}\left( 0\right) $ and
$y_{2}\left( 0\right)
$ via the simple formulas (\ref{y12Ex1}) (at $t=0$); \textit{(ii)} compute $%
y_{1}\left( t\right) $ and $y_{2}\left( t\right) $ from the initial data $%
y_{1}\left( 0\right) $ and $y_{2}\left( 0\right) $ via the,
assumedly
\textit{algebraically} \textit{solvable}, system of evolution equations (\ref%
{ymdot12}) characterizing the time-evolution of these variables; \textit{%
(iii)} the variables $x_{1}\left( t\right) $ and $x_{2}\left(
t\right) $ are
then obtained as the $2$ \textit{zeros} of the, now known, polynomial (\ref%
{p2zt}) (via an \textit{algebraic} operation, indeed one that in
this case
of a polynomial of \textit{second-degree} can be performed \textit{explicitly%
}: note however that this operation yields $2$ \textit{a priori}
indistinguishable functions $x_{n}\left( t\right) $ with $n=1,2$; to
identify which is $x_{1}\left( t\right) $ and which is $x_{2}\left(
t\right) $ these solutions must be followed back---by continuity in
time, from the time $t$ to the initial time $0$---to identify which
one of them corresponds
to the initially assigned data $x_{1}\left( 0\right) $ respectively $%
x_{2}\left( 0\right) $).

The new twist of this approach on which the findings reported in
this paper are based is to assume that the two functions
$f_{m}\left( y_{1},y_{2}\right) $ with $m=1,2$---besides implying
the solvability of the
system (\ref{ymdot12})---feature the additional properties to be \textit{%
polynomial} in their arguments and moreover to satisfy
identically---i. e., for all values of the variable $x$---the
relation

\begin{equation}
x~f_{1}\left( -2x,x^{2}\right) +f_{2}\left( -2x,x^{2}\right) =0~,
\label{Cond1}
\end{equation}%
which clearly implies that the $2$ polynomials%
\begin{equation}
x_{n}f_{1}\left( -x_{1}-x_{2},x_{1}x_{2}\right) +f_{2}\left(
-x_{1}-x_{2},x_{1}x_{2}\right) ~,~~~n=1,2
\end{equation}%
contain \emph{both} the factor $x_{1}-x_{2}$. Therefore this condition (\ref{Cond1}%
) is sufficient to imply that the system of ODEs (\ref{xndot12}) in
fact feature a \textit{polynomial} right-hand side:

\begin{equation}
\dot{x}_{n}=P^{\left( n\right) }\left( x_{1},x_{2}\right)
~,~~~n=1,2~, \label{xndotPol}
\end{equation}%
with $P^{\left( n\right) }\left( x_{1},x_{2}\right) $
\textit{polynomial} in its $2$ arguments.

In the following Section 2 we discuss a rather simple example
manufactured in this manner (hereafter referred to as
\textbf{Example 1}), the equations of motion of which read as
follows:

\textbf{Example 1}:%
\begin{equation}
\dot{x}_{n}=a+b\left[ \left( x_{n}\right) ^{2}-4x_{1}x_{2}-\left(
x_{n+1}\right) ^{2}\right] ~,~~~n=1,2~~~~{mod}\left[ 2\right] ~,
\label{Ex11}
\end{equation}%
with $a$ and $b$ two arbitrary parameters.$\Box$

In Section 3 and its subsections we discuss $3$ other somewhat
analogous models (hereafter referred to respectively as
\textbf{Examples 2,3,4}) obtained via a recent development of the
above approach to identify \textit{algebraically solvable} dynamical
systems, in which the role of the \textit{generic} polynomial
(\ref{PolN}) is however replaced by a polynomial featuring, for all
time, \textit{one double zero }\cite{BC2018}. The equations of
motion characterizing these $3$ dynamical systems read as follows:

\textbf{Example 2}:%
\begin{eqnarray}
\dot{x}_{1} &=&a+b\left[ \left( x_{1}\right) ^{2}+7x_{1}x_{2}+\left(
x_{2}\right) ^{2}\right] ~ \nonumber,
\end{eqnarray}
\begin{eqnarray}
\dot{x}_{2} &=&a+b\left[ 7\left( x_{1}\right)
^{2}+4x_{1}x_{2}-2\left( x_{2}\right) ^{2}\right]   ; \label{Ex21}
\end{eqnarray}

\textbf{Example 3}:%
\begin{equation}
\dot{x}_{n}=x_{n}\left[ a-b\left( x_{1}\right) ^{2}x_{2}\right]
~,~~~n=1,2~ ~;  \label{Ex22}
\end{equation}

\textbf{Example 4}:%
\begin{equation}
\dot{x}_{n}=x_{n}\left[ a+bx_{1}\left( x_{1}+2x_{2}\right) \right]
~,~~~n=1,2~ ~;  \label{Ex23}
\end{equation}%
again, in each of these $3$ cases, with $a$ and $b$ two arbitrary
parameters.

\textbf{Remark 1.1}. Of course in all these examples the presence of
the $2$ \textit{a priori arbitrary} parameters $a$ and $b$ is
somewhat insignificant: indeed, both can clearly be replaced by
\textit{unity }by rescaling the independent variable ($t\Rightarrow
\alpha t$) and the dependent variables ($x_{n}\Rightarrow \beta
x_{n}$) (with obvious appropriate assignments of the parameters
$\alpha $ and $\beta $). Moreover all these examples with an
\textit{arbitrary nonvanishing} value of the parameter $a$ can be
obtained via analogous models with $a=0$ via a simple change of the
independent variable (see below \textbf{Subsection 4.2}). While
models featuring more arbitrary parameters can be derived from these
via a simple change of dependent variables (see below
\textbf{Subsection 4.1}).$~\Box $

Indeed, in \textbf{Section 4} and its subsections we tersely outline
some variants of the examples discussed in \textbf{Sections 2} and
\textbf{3}, thereby enlarging the class of \textit{algebraically
solvable} dynamical systems identifiable via the technique
introduced in this paper. These models might be of interest in
applicative contexts: indeed, dynamical systems of the type
discussed in this paper play a role in an ample variety of such
contexts (say, from population dynamics to chemical reaction to
econometric projections, etc.: you name it). But in this paper we
merely focus on the presentation of the
technique that subtends the identification of this kind of \textit{%
algebraically solvable} dynamical systems characterized by coupled
systems
of $2$ first-order ODEs with polynomial right-hand sides, see (\ref{xndotPol}%
).

Finally \textbf{Section 5} mentions possible future developments of
these findings.

\section{Example 1}

In this \textbf{Section 2} we demonstrate the \textit{algebraically
solvable} character of the dynamical system (\ref{Ex11}).

The starting point of our treatment is the dynamical system (\ref{xnymdot12}%
) with

\begin{equation}
\dot{y}_{1}=\alpha _{0}+\alpha _{1}y_{2}~,~~~\dot{y}_{2}=\beta
_{0}y_{1}+\beta _{1}\left( y_{1}\right) ^{3}~,  \label{Ex11ya}
\end{equation}%
corresponding to (\ref{ymdot12}) with%
\begin{equation}
f_{1}\left( y_{1},y_{2}\right) =\alpha _{0}+\alpha
_{1}y_{2}~,~~~f_{2}\left( y_{1},y_{2}\right) =\beta _{0}y_{1}+\beta
_{1}\left( y_{1}\right) ^{3}~, \label{f12Ex1}
\end{equation}%
where $\alpha _{0},~\alpha _{1},~\beta _{0},~\beta _{1}$ are $4$
\textit{a priori arbitrary} parameters.

These equations of motion clearly imply that the condition
(\ref{Cond1}) is satisfied provided

\begin{equation}
\alpha _{0}=2\beta _{0}~,~~~\alpha _{1}=8\beta _{1}~;
\label{alphabeta11}
\end{equation}%
and it is as well easily seen that there then obtains the system (\ref{Ex11}%
) with $a=-\beta _{0}$, $b=\beta _{1}$, via the insertion of
(\ref{Ex11ya}) in (\ref{xnymdot12}) (of course with $y_{1}=-\left(
x_{1}+x_{2}\right) $ and $y_{2}=x_{1}x_{2}$: see (\ref{y12Ex1})).

On the other hand it is easily seen that the system (\ref{Ex11ya})
is \textit{explicitly solvable}: indeed the equations of motion
(\ref{Ex11ya}) clearly imply the \textit{second-order} ODE (of
Newtonian type: "acceleration equal force")

\label{Ex1ydotdot}
\begin{equation}
\ddot{y}_{1}=\alpha _{1}\left[ \beta _{0}y_{1}+\beta _{1}\left(
y_{1}\right) ^{3}\right] ~,  \label{Ex11yb}
\end{equation}%
namely, via (\ref{alphabeta11}),%
\begin{equation}
\ddot{y}_{1}=8\beta _{1}\left[ \beta _{0}y_{1}+\beta _{1}\left(
y_{1}\right) ^{3}\right] ~.  \label{E11ybb}
\end{equation}%
This second-order ODE---which is of course integrable via two
quadratures---is clearly the Newtonian equation of motion of the
simplest anharmonic oscillator (although, in the \textit{real}
domain, with a force that at large distance pushes the solution away
from the origin). The most direct way to demonstrate the
\textit{algebraically solvable} character of this equation of motion
is to exhibit its solution which---as the interested
reader will easily verify---reads, in terms (for instance) of the \textit{%
first} Jacobian elliptic function $sn(z)$ (see for instance
\cite{HTF2}), as follows:

\begin{equation}
y_{1}\left( t\right) =\mu ~ sn\left(\lambda t+\rho ,k\right) ,
\end{equation}
where $\mu $ and $\lambda $ are determined in terms of the parameter
$k$ as follows:
\begin{equation}
\lambda ^{2}=-\frac{8\beta _{0}\beta _{1}}{1+k^{2}}~,~~~\mu ^{2}=-\frac{%
2\beta _{0}k^{2}}{\beta _{1}\left( 1+k^{2}\right) }~,
\end{equation}%
$\rho $ is determined in terms of the initial datum $y_{1}\left(
0\right) $
as follows,%
\begin{equation}
y_{1}\left( 0\right) =\mu ~sn\left( \rho ,k\right) ~,
\end{equation}%
and the parameter $k$ is determined in terms of the initial data $%
y_{1}\left( 0\right) $ and $y_{2}\left( 0\right) $ as the solution
of the
following algebraic equation%
\begin{equation}
\left\{ \left[ \dot{y}_{1}\left( 0\right) \right] ^{2}-8\beta _{0}\beta _{1}%
\left[ y_{1}\left( 0\right) \right] ^{2}-\left( 2\beta _{1}\right)
^{2}\left[ y_{1}\left( 0\right) \right] ^{4}\right\} \left(
1+k^{2}\right) ^{2}=\left( 4\beta _{0}\right) ^{2}k^{2}~,
\label{Eqk2}
\end{equation}%
where of course (see the first (\ref{Ex11ya}) with (\ref{alphabeta11}))%
\begin{equation}
\dot{y}_{1}\left( 0\right) =2\beta _{0}+8\beta _{1}y_{2}\left(
0\right) ~. \label{ydot0y2}
\end{equation}

And of course, once $y_{1}\left( t\right) $ is known, $y_{2}\left(
t\right) $ is given directly by the first (\ref{Ex11ya}) with
(\ref{alphabeta11}).

\textbf{Remark 2.1}. For given assigned values of $y_{1}\left(
0\right) $
and $y_{2}\left( 0\right) $ (hence $\dot{y}_{1}\left( 0\right) $, see (\ref%
{ydot0y2})), (\ref{Eqk2}) is a \textit{quadratic} equation for
$k^{2}$; the choice of the appropriate solution for $k^{2}$ among
the $2$ solutions of
this elementary equation must of course be made \textit{cum grano salis}. $%
\Box $

\section{Examples 2, 3 and 4}

In the 3 subsections of this \textbf{Section 3} we demonstrate the \textit{%
algebraically solvable} character of the $3$ dynamical systems
(\ref{Ex21}), (\ref{Ex22}) and (\ref{Ex23}).

But let us first summarize some relevant findings of \cite{BC2018}.

Let $p_{3}\left( z;t\right) $ be a time-dependent polynomial of \textit{third%
} degree in its argument $z$ which, for all time, features a
\textit{double pole}:

\begin{equation}
p_{3}\left( z;t\right) =z^{3}+\sum_{m=1}^{3}\left[ y_{m}\left(
t\right) z^{3-m}\right] =\left[ z-x_{1}\left( t\right) \right]
^{2}\left[ z-x_{2}\left( t\right) \right] ~.  \label{p3zt}
\end{equation}%
This of course implies that its $3$ \textit{coefficients}
$y_{m}\left(
t\right) $ are expressed as follows in terms of the \textit{double zero} $%
x_{1}\left( t\right) $ and the \textit{zero }(of unit multiplicity) $%
x_{2}\left( t\right) $:%
\begin{equation}
y_{1}=-\left( 2x_{1}+x_{2}\right) ~,~~~y_{2}=x_{1}\left(
x_{1}+2x_{2}\right) ~,~~~y_{3}=-\left( x_{1}\right) ^{2}x_{2}~;
\label{yyy}
\end{equation}%
and correspondingly that the $3$ coefficients $y_{m}\left( t\right)
$ are, for all time, related to each other by the (single) condition
implied by the simultaneous vanishing at $z=x_{1}\left( t\right) $
of both $p_{3}\left(
z;t\right) $ and its $z$-derivative $p_{3,z}\left( z;t\right) $:%
\begin{equation}
p_{3}\left( x_{1};t\right) =\left[ x_{1}\left( t\right) \right]
^{3}+\sum_{m=1}^{3}\left\{ y_{m}\left( t\right) \left[ x_{1}\left( t\right) %
\right] ^{3-m}\right\} =0~,  \label{p3x1}
\end{equation}%
\begin{equation}
p_{3,z}\left( x_{1};t\right) =3\left[ x_{1}\left( t\right) \right]
^{2}+2y_{1}\left( t\right) x_{1}\left( t\right) +y_{2}\left(
t\right) =0~. \label{p3zx1}
\end{equation}%
In an analogous manner (see the treatment in \textbf{Section 2}, and if need be \cite%
{BC2018}) it is possible to obtain the following $3$ pairs of
formulas
(analogous to, but of course somewhat different from, the formulas (\ref%
{xnymdot12})):

\label{x12dotEx}
\begin{equation}
\dot{x}_{1}=-\frac{2~x_{1}~\dot{y}_{1}+\dot{y}_{2}}{2~\left(
x_{1}-x_{2}\right) }~,~~~\dot{x}_{2}=\frac{\left( x_{1}+x_{2}\right) ~\dot{y}%
_{1}+\dot{y}_{2}}{x_{1}-x_{2}}~;  \label{x12dotEx2}
\end{equation}%
\begin{equation}
\dot{x}_{1}=-\frac{\left( x_{1}\right) ^{2}~\dot{y}_{1}-\dot{y}_{3}}{%
2~x_{1}~\left( x_{1}-x_{2}\right) }~,~~~\dot{x}_{2}=\frac{x_{1}~x_{2}~\dot{y}%
_{1}-\dot{y}_{3}}{x_{1}~\left( x_{1}-x_{2}\right) }~;
\label{x12dotEx3}
\end{equation}%
\begin{equation}
\dot{x}_{1}=\frac{x_{1}~\dot{y}_{2}+2~\dot{y}_{3}}{2~x_{1}~\left(
x_{1}-x_{2}\right)
}~,~~~\dot{x}_{2}=-\frac{x_{1}~x_{2}~\dot{y}_{2}+\left(
x_{1}+x_{2}\right) ~\dot{y}_{3}}{\left( x_{1}\right) ^{2}~\left(
x_{1}-x_{2}\right) }~.  \label{x12dotEx4}
\end{equation}

It is then clear---in close analogy to the treatment described above
(\textbf{see Section 1})---that each of these $3$ pairs of formulas
opens the way to the identification of \textit{algebraically
solvable} dynamical systems
involving the $2$ dependent variables $x_{1}\left( t\right) $ and $%
x_{2}\left( t\right) $: as separately discussed in the following $3$
subsections.

\subsection{Example 2}

In this \textbf{Subsection 3.1} we demonstrate the
\textit{algebraically solvable} character of the dynamical systems
(\ref{Ex21}).

Now the starting point of our treatment is---instead of the system (\ref%
{xnymdot12})--- the slightly different system (\ref{x12dotEx2}).
Clearly
this system is \textit{solvable by algebraic operations} if the quantities $%
y_{1}\left( t\right) $ and $y_{2}\left( t\right) $ satisfy the system (\ref%
{ymdot12}) and this system is itself \textit{solvable.} Then the
system satisfied by the variables $x_{1}\left( t\right) $ and
$x_{2}\left( t\right)
$---obtained by replacing, in the right hand side of (\ref{Ex21}), $\dot{y}%
_{1}$ and $\dot{y}_{2}$ via the equations of motion
(\ref{ymdot12})---reads

\begin{eqnarray}
\dot{x}_{1} &=&-\frac{2x_{1}f_{1}\left( -2x_{1}-x_{2},\left(
x_{1}\right) ^{2}+2x_{1}x_{2}\right) +f_{2}\left(
-2x_{1}-x_{2},\left( x_{1}\right)
^{2}+2x_{1}x_{2}\right) }{2~\left( x_{1}-x_{2}\right) }~,  \nonumber \\
\dot{x}_{2} &=&\left( x_{1}-x_{2}\right) ^{-1}\left[ \left(
x_{1}+x_{2}\right) f_{1}\left( -2x_{1}-x_{2},\left( x_{1}\right)
^{2}+2x_{1}x_{2}\right) \right.  \nonumber \\
&&\left. +f_{2}\left( -2x_{1}-x_{2},\left( x_{1}\right)
^{2}+2x_{1}x_{2}\right) \right] ~,  \label{Ex2b}
\end{eqnarray}%
corresponding now to the assignment (\ref{yyy}) (instead of
(\ref{y12Ex1}))
of $y_{1}\left( t\right) $ and $y_{2}\left( t\right) $ in terms of $%
x_{1}\left( t\right) \ $and $x_{2}\left( t\right) $.

It is now clear that the conditions on the $2$ functions
$f_{1}\left( y_{1},y_{2}\right) $ and $f_{2}\left(
y_{1},y_{2}\right) $ which are
sufficient to guarantee that the right-hand side of the equations of motion (%
\ref{Ex2b}) be \textit{polynomial} in the $2$ dependent variables $%
x_{1}\left( t\right) $ and $x_{2}\left( t\right) $ are that these
$2$ functions $f_{1}\left( y_{1},y_{2}\right) $ and $f_{2}\left(
y_{1},y_{2}\right) $ be themselves \textit{polynomial} in their $2$
variables $y_{1}$ and $y_{2}$ and moreover that there hold
identically---i. e., for all values of $x$---the relation
\begin{equation}
2xf_{1}\left( -3x,3x^{2}\right) +f_{2}\left( -3x,3x^{2}\right) =0~.
\label{Cond2}
\end{equation}

We now assume that the time-evolution of the $2$ quantities
$y_{1}\left( t\right) $ and $y_{2}\left( t\right) $ be again
characterized by the equations of motion (\ref{Ex11ya})---the
\textit{solvable} character of
which has been pointed out in \textbf{Section 2}---hence by the assignments (\ref%
{f12Ex1}) of the two functions $f_{1}\left( y_{1},y_{2}\right) $ and $%
f_{2}\left( y_{1},y_{2}\right) $. It is then easily seen that the condition (%
\ref{Cond2}) entails now the relations%
\begin{equation}
\alpha _{0}=\frac{3\beta _{0}}{2}~,~~~\alpha _{1}=\frac{9\beta
_{1}}{2} \label{alphabeta2}
\end{equation}%
(instead of (\ref{alphabeta11})).

It is then easily seen that the corresponding dynamical system
satisfied by the coordinates $x_{1}\left( t\right) $ and
$x_{2}\left( t\right) $ is just the system of $2$ ODEs (\ref{Ex21})
(with $a=-\beta _{0}/2,$ $b=-\beta _{1}/2 $).

There remains to report---from \cite{BC2018}---how to obtain from
the variables $y_{1}\left( t\right) $ and $y_{2}\left( t\right) $ the variables $%
x_{1}\left( t\right) $ and $x_{2}\left( t\right) .$ The variable $%
x_{1}\left( t\right) $ is that one of the $2$ roots of the---of
course
\textit{explicitly solvable}---second-degree polynomial equation in $x$%
\begin{equation}
3x^{2}+2y_{1}\left( t\right) x+y_{2}\left( t\right) =0
\end{equation}%
(see (\ref{p3zx1})) which, by continuity in $t,$ corresponds at
$t=0$ to the initially assigned datum $x_{1}\left( 0\right) .$ While
$x_{2}\left( t\right) $ is then given by the formula
\begin{equation}
x_{2}\left( t\right) =-y_{1}\left( t\right) -2x_{1}\left( t\right)
\end{equation}%
(see the first of the 3 formulas (\ref{yyy})).

\subsection{Example 3}

In this \textbf{Subsection 3.2} we demonstrate the
\textit{algebraically solvable} character of the dynamical systems
(\ref{Ex22}).

Now the starting point of our treatment is the system of $2$ coupled ODEs (%
\ref{x12dotEx3}). Clearly this system is \textit{solvable by
algebraic operations} if the quantities $y_{1}\left( t\right) $ and
$y_{3}\left(
t\right) $ satisfy the system%
\begin{equation}
\dot{y}_{1}=f_{1}\left( y_{1},y_{3}\right)
~,~~~\dot{y}_{3}=f_{3}\left( y_{1},y_{3}\right) ~,  \label{ymdot13}
\end{equation}%
and this system is itself \textit{solvable} (of course for an
appropriate
assignment of the $2$ functions $f_{1}\left( y_{1},y_{3}\right) $ and $%
f_{3}\left( y_{1},y_{3}\right) $)\textit{.} Then the system
satisfied by the variables $x_{1}\left( t\right) $ and $x_{2}\left(
t\right) $---obtained by
replacing, in the right-hand side of (\ref{x12dotEx3}), $\dot{y}_{1}$ and $%
\dot{y}_{3}$ via these equations of motion (\ref{ymdot13})---reads
\begin{eqnarray}
\dot{x}_{1} &=&\frac{-\left( x_{1}\right) ^{2}f_{1}\left(
-2x_{1}-x_{2},-\left( x_{1}\right) ^{2}x_{2}\right) +f_{3}\left(
-2x_{1}-x_{2},-\left( x_{1}\right) ^{2}x_{2}\right) }{2x_{1}\left(
x_{1}-x_{2}\right) }~,  \nonumber \\
\dot{x}_{2} &=&\left[ x_{1}\left( x_{1}-x_{2}\right) \right]
^{-1}\left[ x_{1}x_{2}f_{1}\left( -2x_{1}-x_{2},\left( x_{1}\right)
^{2}+2x_{1}x_{2}\right) \right.  \nonumber \\
&&\left. -f_{3}\left( -2x_{1}-x_{2},\left( x_{1}\right)
^{2}+2x_{1}x_{2}\right) \right] ~,  \label{Ex3}
\end{eqnarray}%
corresponding to the assignment (\ref{yyy}) of $y_{1}\left( t\right) $ and $%
y_{3}\left( t\right) $ in terms of $x_{1}\left( t\right) \ $and
$x_{2}\left( t\right) $; and it is easily seen that sufficient
conditions to guarantee that this become a system of $2$ ODEs
featuring in their right-hand sides a \textit{polynomial} dependence
on the $2$ dependent variables $x_{1}\left(
t\right) \ $and $x_{2}\left( t\right) $ are that these $2$ functions $%
f_{1}\left( y_{1},y_{2}\right) $ and $f_{3}\left( y_{1},y_{2}\right)
$ be themselves \textit{polynomial} in their $2$ variables $y_{1}$
and $y_{3}$
and moreover that there hold identically---i. e., for all values of $x$%
---the relation%
\begin{equation}
\frac{x^{2}f_{1}\left( -3x,-x^{3}\right) -f_{3}\left( -3x,-x^{3}\right) }{x}%
=0~.  \label{Cond3}
\end{equation}

Let us now assume that the two functions $f_{1}\left(
y_{1},y_{3}\right) $ and $f_{3}\left( y_{1},y_{3}\right) $ read as
follows:

\begin{equation}
f_{1}\left( y_{1},y_{3}\right) =y_{1}\left( \alpha _{1}+\alpha
_{2}y_{3}\right) ~,~~~f_{3}\left( y_{1},y_{3}\right) =y_{3}\left(
\beta _{1}+\beta _{2}y_{3}\right) ~,
\end{equation}%
so that the system (\ref{ymdot13}) reads%
\begin{equation}
\dot{y}_{1}=y_{1}\left( \alpha _{1}+\alpha _{2}y_{3}\right) ~,~~~\dot{y}%
_{3}=y_{3}\left( \beta _{1}+\beta _{2}y_{3}\right) ~.  \label{ydot3}
\end{equation}

Here the $4$ parameters $\alpha _{1},\alpha _{2},\beta _{1},\beta
_{2}$ are \textit{a priori} arbitrary, but clearly to satisfy
(\ref{Cond3}) it is necessary and sufficient that (as we hereafter
assume, in this \textbf{Subsection 3.2})

\begin{equation}
\beta _{1}=3\alpha _{1}~,~~~\beta _{2}=3\alpha _{2}~.
\label{betaalpha3}
\end{equation}%
It is then a matter of trivial algebra to verify that the
corresponding system of $2$ ODEs for the $2$ dependent variables
$x_{1}\left( t\right) $
and $x_{2}\left( t\right) $ is just (\ref{Ex22}), with $a=\alpha _{1}$, $%
b=-\alpha _{2}$.

It is on the other hand easily seen that the system (\ref{ydot3}) is \textit{%
explicitly solvable}: by firstly integrating by a quadrature the ODE
satisfied by the dependent variable $y_{3}\left( t\right) ,$ and by
then
integrating the \textit{linear} ODE satisfied by the dependent variable $%
y_{1}\left( t\right) $. There results the following neat expressions of the $%
2$ variables $y_{1}\left( t\right) $ and $y_{3}\left( t\right) $:

\label{y1y3Ex3}
\begin{equation}
y_{1}\left( t\right) =y_{1}\left( 0\right) \varphi \left( t\right)
~,~~~y_{3}\left( t\right) =y_{3}\left( 0\right) \left[ \varphi
\left( t\right) \right] ^{3}~,  \label{y1y3}
\end{equation}%

\begin{equation}
\varphi \left( t\right) =\exp(at)\left\{ 1-\left( \frac{b}{a}\right)
y_{3}\left( 0\right) [1-\exp (3at)]\right\} ^{-1/3}~. \label{phi}
\end{equation}

The subsequent computation of the $2$ dependent variables
$x_{1}\left( t\right) $ and $x_{2}\left( t\right) $ from the $2$
quantities $y_{1}\left( t\right) $ and $y_{3}\left( t\right) $ can
then be easily performed: it involves the \textit{algebraic}
operation of solving a \textit{cubic} equation (a task which can
actually be performed explicitly), as the interested reader will
easily ascertain (or, if need be, see \cite{BC2018}).

\textbf{Remark 3.2.1}. If the parameter $a$ is \textit{imaginary}---$a=%
\mathbf{i}\omega $ (with, here and hereafter, $\mathbf{i}$ the \textit{%
imaginary} \textit{unit}, so that $\mathbf{i}^{2}=-1$) and $\omega $ is \textit{%
real} and \textit{nonvanishing}, $\omega \neq 0$---both
\textit{coefficients} $y_{1}\left( t\right) $ and $y_{3}\left(
t\right) $ are clearly periodic with period $T=2\pi /\left\vert
\omega \right\vert $, see (\ref{y1y3}) with (\ref{phi}): actually
$y_{3}\left( t\right) $ is clearly periodic with period $T/3$; while
$y_{1}\left( t\right) $ is certainly periodic with period $T$
but---depending on the value of the initial datum $y_{3}\left( 0\right) $%
---it might also be periodic with period $T/3$. Hence the $2$ coordinates $%
x_{1}\left( t\right) $ and $x_{2}\left( t\right) $ are themselves
periodic with period $T$ (or possibly a small integer multiple of $T;$ see \cite%
{GS2005} \cite{C2008}).$\Box$

\subsection{Example 4}

In this \textbf{Subsection 3.3} we demonstrate the
\textit{algebraically solvable} character of the dynamical systems
(\ref{Ex23}).

Now the starting point of our treatment is system (\ref{x12dotEx4}).
Clearly
this system is \textit{solvable by algebraic operations} if the quantities $%
y_{2}\left( t\right) $ and $y_{3}\left( t\right) $ satisfy the
system

\begin{equation}
\dot{y}_{2}=f_{2}\left( y_{2},y_{3}\right)
~,~~~\dot{y}_{3}=f_{3}\left( y_{2},y_{3}\right) ~,  \label{ymdot14}
\end{equation}%
and this system is itself \textit{solvable} (of course for an
appropriate
assignment of the $2$ functions $f_{2}\left( y_{1},y_{3}\right) $ and $%
f_{3}\left( y_{1},y_{3}\right) $)\textit{.} Then the system
satisfied by the variables $x_{1}\left( t\right) $ and $x_{2}\left(
t\right) $---obtained by
replacing, in the right-hand side of (\ref{x12dotEx4}), $\dot{y}_{2}$ and $%
\dot{y}_{3}$ via these equations of motion (\ref{ymdot14})---reads
\begin{eqnarray}
\dot{x}_{1} &=&\frac{x_{1}f_{2}\left( x_{1}\left(
x_{1}+2x_{2}\right) ,-\left( x_{1}\right) ^{2}x_{2}\right)
+2f_{3}\left( x_{1}\left( x_{1}+2x_{2}\right) ,-\left( x_{1}\right)
^{2}x_{2}\right) }{2x_{1}\left(
x_{1}-x_{2}\right) }~,  \nonumber \\
\dot{x}_{2} &=&-\left( x_{1}\right) ^{-2}\left( x_{1}-x_{2}\right)
^{-1}\left\{ x_{1}x_{2}f_{2}\left( x_{1}\left( x_{1}+2x_{2}\right)
,-\left(
x_{1}\right) ^{2}x_{2}\right) \right.  \nonumber \\
&&\left. +\left( x_{1}+x_{2}\right) f_{3}\left( x_{1}\left(
x_{1}+2x_{2}\right) ,-\left( x_{1}\right) ^{2}x_{2}\right) \right\}
~,
\end{eqnarray}%
corresponding to the assignment (\ref{yyy}) of $y_{2}\left( t\right) $ and $%
y_{3}\left( t\right) $ in terms of $x_{1}\left( t\right) \ $and
$x_{2}\left( t\right) $; and it is easily seen that sufficient
conditions to guarantee that this become a system of $2$ ODEs
featuring in their right-hand sides a \textit{polynomial} dependence
on the $2$ dependent variables $x_{1}\left(
t\right) \ $and $x_{2}\left( t\right) $ are that these $2$ functions $%
f_{2}\left( y_{1},y_{2}\right) $ and $f_{3}\left( y_{1},y_{2}\right)
$ be themselves \textit{polynomial} in their $2$ variables $y_{2}$
and $y_{3}$
and moreover that there hold identically---i. e., for all values of $x$%
---the relation
\begin{equation}
\frac{xf_{2}\left( 3\left( x\right) ^{2},-\left( x\right)
^{3}\right) +2f_{3}\left( 3\left( x\right) ^{2},-\left( x\right)
^{3}\right) }{x}=0~. \label{Cond4}
\end{equation}

Let us now assume that the two functions $f_{2}\left(
y_{1},y_{3}\right) $ and $f_{3}\left( y_{1},y_{3}\right) $ read as
follows:

\begin{equation}
f_{2}\left( y_{2},y_{3}\right) =y_{2}\left( \alpha _{1}+\alpha
_{2}y_{2}\right) ~,~~~f_{3}\left( y_{2},y_{3}\right) =y_{3}\left(
\beta _{1}+\beta _{2}y_{2}\right) ~,
\end{equation}%
so that the system (\ref{ymdot13}) reads%
\begin{equation}
\dot{y}_{2}=y_{2}\left( \alpha _{1}+\alpha _{2}y_{2}\right) ~,~~~\dot{y}%
_{3}=y_{3}\left( \beta _{1}+\beta _{2}y_{2}\right) ~.  \label{ydot4}
\end{equation}

Here the $4$ parameters $\alpha _{1},\alpha _{2},\beta _{1},\beta
_{2}$ are \textit{a priori} arbitrary, but clearly to satisfy
(\ref{Cond4}) it is necessary and sufficient that (as we hereafter
assume, in this \textbf{Subsection 3.3})

\begin{equation}
\beta _{1}=\frac{3\alpha _{1}}{2}~,~~~\beta _{2}=\frac{3\alpha
_{2}}{2}~.
\end{equation}%
It is then a matter of trivial algebra to verify that the
corresponding system of $2$ ODEs for the $2$ dependent variables
$x_{1}\left( t\right) $
and $x_{2}\left( t\right) $ is just (\ref{Ex23}), with $a=\alpha _{1}/2$, $%
b=\alpha _{2}/2$.

It is on the other hand easily seen that the system (\ref{ydot4}) is \textit{%
explicitly solvable}: it is indeed, up to simple notational changes,
identical to the system (\ref{ydot3}) discussed in the preceding
\textbf{Subsection 3.2}.

And the subsequent computation of the $2$ dependent variables
$x_{1}\left( t\right) $ and $x_{2}\left( t\right) $ from the $2$
quantities $y_{2}\left( t\right) $ and $y_{3}\left( t\right) $ can
as well be easily performed: it involves again the
\textit{algebraic} operation of solving a \textit{cubic} equation (a
task which can actually be performed explicitly), as the interested
reader will easily ascertain (or, if need be, see \cite{BC2018}).

\section{Variants}

In this \textbf{Section 4} and its subsections we tersely outline
some interesting variants of the \textit{algebraically solvable}
models discussed above, which might be of interest for possible
utilizations of these findings in applicative contexts.

\subsection{First variant}

Each of the $4$ dynamical systems identified above as
\textit{algebraically
solvable}---see (\ref{Ex11}), (\ref{Ex21}), (\ref{Ex22}), (\ref{Ex23}%
)---features only $2$ arbitrary parameters, $a$ and $b$. Systems
featuring more free parameters can of course be obtained from these
via the trivial change of dependent variables

\label{transu}
\begin{equation}
x_{1}\left( t\right) =u_{10}+u_{11}\xi _{1}\left( t\right)
+u_{12}\xi _{2}\left( t\right) ~,~~~x_{2}\left( t\right)
=u_{20}+u_{21}\xi _{1}\left( t\right) +u_{22}\xi _{2}\left( t\right)
~,\label{transu1}
\end{equation}%
featuring the $6$ parameters $u_{n\ell }$, $n=1,2$, $\ell =0,1,2$.
This
change of variables is easily inverted:%
\begin{eqnarray}
\xi _{1}\left( t\right) &=&u^{-1}\left\{ u_{22}\left[ x_{1}\left(
t\right) -u_{10}\right] -u_{12}\left[ x_{2}\left( t\right)
-u_{20}\right] \right\} ~,
\nonumber \\
\xi _{2}\left( t\right) &=&u^{-1}\left\{ u_{21}\left[ x_{1}\left(
t\right) -u_{10}\right] -u_{11}\left[ x_{2}\left( t\right)
-u_{20}\right] \right\} ~,
\end{eqnarray}%
where, here and hereafter,%
\begin{equation}
u=u_{11}u_{22}-u_{12}u_{21}~.  \label{u}
\end{equation}%
Clearly the properties of \textit{algebraic solvability} are not
affected, although the relevant formulas become marginally more
complicated, requiring the solution of some (\textit{purely
algebraic}) equations. On the other
hand the new systems of $2$ ODEs satisfied by the new dependent variables $%
\xi _{1}\left( t\right) $ and $\xi _{2}\left( t\right) $ feature now
several
more free parameters. For instance for \textbf{Example 1} the equations that replace (%
\ref{Ex11}) read as follows:

\begin{equation}
\dot{\xi}_{n}=A_{n}+B_{n1}\xi _{1}+B_{n2}\xi _{2}+C_{n1}\left( \xi
_{1}\right) ^{2}+C_{n2}\left( \xi _{2}\right) ^{2}+C_{n3}\xi _{1}\xi
_{2}~, \label{ksindot}
\end{equation}%
\begin{equation}
A_{1}=u^{-1}\left\{ \left( u_{22}-u_{12}\right) \left(
a-4bu_{10}u_{20}\right) +b\left( u_{22}+u_{12}\right) \left[ \left(
u_{10}\right) ^{2}-\left( u_{20}\right) ^{2}\right] \right\} ~,
\end{equation}%
\begin{equation}
A_{2}=u^{-1}\left\{ \left( u_{21}-u_{11}\right) \left(
a-4bu_{10}u_{20}\right) +b\left( u_{21}+u_{11}\right) \left[ \left(
u_{10}\right) ^{2}-\left( u_{20}\right) ^{2}\right] \right\} ~,
\end{equation}%
\begin{eqnarray}
B_{1n}=2bu^{-1}\left[ 2\left( u_{12}-u_{22}\right) \left(
u_{10}u_{2n}+u_{20}u_{1n}\right) \right. &&  \nonumber \\
\left. +\left( u_{22}+u_{12}\right) \left(
u_{10}u_{1n}-u_{20}u_{2n}\right) \right] ~,~~~n=1,2~, &&
\end{eqnarray}%
\begin{eqnarray}
B_{2n}=2bu^{-1}\left[ 2\left( u_{11}-u_{21}\right) \left(
u_{10}u_{2n}+u_{20}u_{1n}\right) \right. &&  \nonumber \\
\left. +\left( u_{21}+u_{11}\right) \left(
u_{10}u_{1n}-u_{20}u_{2n}\right) \right] ~,~~~n=1,2~, &&
\end{eqnarray}%
\begin{equation}
C_{1n}=bu^{-1}\left\{ 4\left( u_{12}-u_{22}\right)
u_{1n}u_{2n}+\left( u_{22}+u_{12}\right) \left[ \left( u_{1n}\right)
^{2}-\left( u_{2n}\right) ^{2}\right] \right\} ~,~~~n=1,2~,
\end{equation}%
\begin{equation}
C_{2n}=bu^{-1}\left\{ 4\left( u_{11}-u_{21}\right)
u_{1n}u_{2n}+\left( u_{21}+u_{11}\right) \left[ \left( u_{1n}\right)
^{2}-\left( u_{2n}\right) ^{2}\right] \right\} ~,~~~n=1,2~,
\end{equation}%
\begin{equation}
C_{13}=2bu^{-1}\left[ 2\left( u_{12}-u_{22}\right) \left(
u_{11}u_{22}+u_{12}u_{21}\right) +\left( u_{22}+u_{12}\right) \left(
u_{11}u_{12}-u_{21}u_{22}\right) \right] ~,
\end{equation}%
\begin{equation}
C_{23}=2bu^{-1}\left[ 2\left( u_{11}-u_{21}\right) \left(
u_{11}u_{22}+u_{12}u_{21}\right) +\left( u_{21}+u_{11}\right) \left(
u_{11}u_{12}-u_{21}u_{22}\right) \right] ~.
\end{equation}

Note that if $a=$ $u_{10}=u_{20}=0$ then $A_{n}=B_{nm}=0$ and the
equations of motion (\ref{ksindot}) have homogeneous right-hand
sides (of degree $2$) featuring only the $6$ coefficients $C_{n\ell
}$ with $n=1,2$ and $\ell =1,2,3,$ expressed in terms of the $5$
\textit{arbitrary} parameters $b$ and $u_{nm}$ with $n$ and $m$
taking the values $1$ and $2$.

It is left to the interested reader to obtain analogous formulas for
\textbf{Examples 2, 3, 4}.

\subsection{Second variant}

If the (\textit{autonomous})\textit{\ }system of $2$ coupled ODEs

\label{scaling}
\begin{equation}
\dot{x}_{n}=f_{n}\left( x_{1},x_{2}\right) ~,~~~n=1,2~,
\label{xndot42}
\end{equation}%
features \textit{homogeneous} functions $f_{n}\left(
x_{1},x_{2}\right) $
satisfying the scaling property%
\begin{equation}
f_{n}\left( cx_{1},cx_{2}\right) =c^{p}f_{n}\left(
x_{1},x_{2}\right) ~,~~~n=1,2~,~~~p\neq 1  \label{c}
\end{equation}%
(where $c$ is an arbitrary parameter), then by setting

\label{Eqwn}
\begin{equation}
w_{n}\left( t\right) =\exp \left( \frac{\alpha t}{p-1}\right)
~x_{n}\left( \tau \left( t\right) \right) ~,~~~\tau \left( t\right)
=\frac{\exp \left( \alpha t\right) -1}{\alpha }~,~~~n=1,2~,
\label{wu}
\end{equation}%
one gets for the new dependent variables $w_{n}\left( t\right) $ the new (%
\textit{autonomous}!) system
\begin{equation}
\dot{w}_{n}=\frac{\alpha }{p-1}w_{n}+f_{n}\left( w_{1},w_{2}\right)
~,~~~n=1,2~.  \label{wndot}
\end{equation}%
Then---if the original system (\ref{xndot42}) is
\textit{algebraically solvable}---the solutions $w_{n}\left(
t\right) $ of this system satisfy interesting properties: in
particular, if $\alpha =\mathbf{i}\omega $ is
\textit{imaginary}---with $\omega $ a \textit{nonvanishing}
\textit{real}
parameter and\textit{\ }$p$ a \textit{real rational} number---then \textit{%
all} solutions of these systems (\ref{wndot}) are \textit{completely periodic%
} with some \textit{rational integer} multiple of the basic period
$T=2\pi /\left\vert \omega \right\vert $ (\textit{isochrony}!). For
more details on the transformation (\ref{wu}) and its implications
see \cite{C2008} and references therein.

Note that the $4$ dynamical systems of \textbf{Examples 1,2,3,4} belong to the class (%
\ref{c}) if the parameter $a$ vanishes, $a=0$: with $p=2$ in the
cases of \textbf{Examples 1} and \textbf{2} (see (\ref{Ex11}) and
(\ref{Ex21})), with $p=4$ in the
case of \textbf{Example 3} and $p=3$ in the case of \textbf{Example 4} (see (\ref{Ex22}) and (%
\ref{Ex23})); and these properties continue to hold after the
generalization described in the preceding \textbf{Subsection 4.1},
provided the parameters $u_{10}$ and $u_{20}$ vanish, $u_{10}=u_{20}=0$ (see (\ref{transu1}%
)).

\subsection{Third variant}

Let us note that the dynamical systems detailed in the
\textbf{Examples} reported
above can be reformulated as describing the evolution of \textit{real} $2$%
-vectors $\vec{r}_{n}\left( t\right) $ lying in a (\textit{real})
plane. Indeed set

\label{Trans43}
\begin{equation}
\vec{r}_{n}\left(t\right)\equiv\left(Re\left[x_{n}\left(t\right)
\right] ,~ Im\left[x_{n}\left(t\right)\right]\right)
~,~~~n=1,2~,\label{Trans44}
\end{equation}%

\begin{equation}
\vec{a}\equiv \left( Re\left[ a\right] ,Im\left[ a\right] \right)
~,~~~\vec{b}\equiv \left( Re\left[ b\right] ,-Im\left[ b\right]
\right) ~.\label{Trans45}
\end{equation}%

Then---as the diligent reader will easily verify---the version of (\ref{Ex11}%
) yielded by this notational change reads as follows:

\begin{eqnarray}
\dot{{\vec{r}}_{n}}=\vec{a}+2\vec{r}_{n}\left[ \vec{b}\cdot \left(
\vec{r}_{n}-2\vec{r}_{n+1}\right) \right] -2\vec{r}_{n+1}\left[
\vec{b}\cdot
\left( \vec{r}_{n+1}+2\vec{r}_{n}\right) \right] &&  \nonumber \\
+\vec{b}\left[ \left( r_{n+1}\right) ^{2}-\left( r_{n}\right)
^{2}+4\left(
\vec{r}_{n}\cdot \vec{r}_{n+1}\right) \right] ~,~~~n=1,2~~~mod\left[ 2%
\right] ~. &&
\end{eqnarray}%
Here of course the dot among two vectors denotes the standard scalar
product, and $\left( r_{n}\right) ^{2}\equiv \vec{r}_{n}\cdot
\vec{r}_{n}$. Note the \textit{covariant} character of these
equations.

The interested reader will have no difficulty to reformulate in an
analogous manner the equations of motion (\ref{Ex21}) of
\textbf{Example 2}; and analogous reformulations of the equations of
motion of \textbf{Examples 3} and \textbf{4} are also
possible (hint: before applying the same procedure as indicated above, see (%
\ref{Trans44}) and (%
\ref{Trans45}), replace $b$ with $b^{3}$ in (\ref{Ex22}), and $b$ with $%
b^{2} $ in (\ref{Ex23})).

\section{Outlook}

The literature on the simple kind of dynamical systems treated in
this paper is of course vast; see for instance \cite{HCF1993},
\cite{N2005} and standard compilations of solvable ODEs such as
\cite{PZ2018}. But it seems to us that---in spite of their
simplicity---the findings reported in this paper (including their
variants mentioned in \textbf{Section 4}) are new.

Further applications of the approach described in this paper are of
course also possible: for instance by exploiting the extension of
the results of
\cite{BC2018} to time-dependent polynomials featuring zeros of \textit{%
arbitrary} multiplicity (see some progress made in this direction by
Oksana Bihun's recent paper \cite{B2018}; we report additional
progress in \cite{CP2018})); or by exploiting the extensions of the
fundamental results---such as (\ref{xnymdot})---on which the
findings reported in this paper are based, from polynomials to
rational functions \cite{C2018a}.

And of course extensions of the approach of this paper to systems of
higher-order ODEs (including in particular second-order ODEs of
Newtonian type:\ "accelerations equal forces"), to PDEs, to
discrete-time evolutions (see \cite{C2018} and \cite{FCP2019})
deserve further investigations.

\section{Acknowledgements}

FP likes to thank the Physics Department of the University of Rome\
"La Sapienza" for the hospitality from March to July 2018 (during
her sabbatical), when the results reported in this paper were
obtained.

\end{document}